\documentclass[aps,pre,10pt,twocolumn,floats,floatfix,showkeys,showpacs,superscriptaddress]{revtex4}
\usepackage{graphicx,amssymb,amsmath}
\usepackage[latin1]{inputenc}
\usepackage{color}
\begin{document}

\title{Exact detection of direct links in networks of interacting dynamical units}
\author{Nicol{\'a}s Rubido}
\affiliation{Institute for Complex Systems and Mathematical Biology, University of Aberdeen,
	     King's College, AB24 3UE Aberdeen, UK}
\email{n.rubido.obrer@abdn.ac.uk}
\affiliation{Instituto de F\'{i}sica, Facultad de Ciencias, Universidad de la Rep\'{u}blica,
	     Igu\'{a} 4225, Montevideo, 11200, Uruguay}
\email{nrubido@fisica.edu.uy}
\author{Arturo C. Mart{\'i}}
\affiliation{Instituto de F\'{i}sica, Facultad de Ciencias, Universidad de la Rep\'{u}blica,
	     Igu\'{a} 4225, Montevideo, 11200, Uruguay}
\author{Ezequiel Bianco-Mart{\'i}nez}
\affiliation{Institute for Complex Systems and Mathematical Biology, University of Aberdeen,
	     King's College, AB24 3UE Aberdeen, UK}
\author{Celso Grebogi}
\affiliation{Institute for Complex Systems and Mathematical Biology, University of Aberdeen,
	     King's College, AB24 3UE Aberdeen, UK}
\author{Murilo S. Baptista}
\affiliation{Institute for Complex Systems and Mathematical Biology, University of Aberdeen,
	     King's College, AB24 3UE Aberdeen, UK}
\author{Cristina Masoller}
\affiliation{Departament de F\'isica i Enginyeria Nuclear, Universitat Polit\'ecnica de
	     Catalunya, Colom 11, E-08222 Terrassa, Barcelona, Spain}
\date{\today}
\begin{abstract}
The inference of an underlying network topology from local observations of a complex system
composed of interacting units is usually attempted by using statistical similarity measures,
such as Cross-Correlation (CC) and Mutual Information (MI). The possible existence of a
direct link between different units is, however, hindered within the time-series measurements.
Here we show that, for the class of systems studied, when an abrupt change in the
ordered set of CC or MI values exists, it is possible to infer, without errors, the
underlying network topology from the time-series measurements, even in the presence of
observational noise, non-identical units, and coupling heterogeneity. We find that a
necessary condition for the discontinuity to occur is that the dynamics of the coupled
units is partially coherent, i.e., neither complete disorder nor globally synchronous
patterns are present. We critically compare the inference methods based on CC and MI, in
terms of how effective, robust, and reliable they are, and conclude that, in general, MI
outperforms CC in robustness and reliability. Our findings could be relevant for the
construction and interpretation of functional networks, such as those constructed from
brain or climate data.
\end{abstract}
\maketitle
 \section{Introduction}
Inferring the underlying topology of a complex system from observed data is currently the
object of intense research. However, the limits for the exact inference of direct links in
real-world systems composed by interacting dynamical units are still not fully understood.
Understanding this limitations is often crucial in many applications in social and natural
sciences. In order to infer the underlying network, usually, the observed data comes from
time-series recorded at the different units. Then, a direct link between units is assumed
depending on how interdependent these observations are. For example, by recording the
activity of different brain regions, one wishes to infer which are the relevant structural
or functional brain connections by comparing similarity patterns \cite{chavez_pre_2008,
Bullmore_review_2009,bjorn_2009}. In general, the outcome is a complex network
\cite{boccaletti_phys_rep_2006, arenas_phys_rep_2008} that interconnects the individual
units and allows for a better understanding of the overall system behavior.

The main statistical tools used to determine the interdependence of the units have been
the Cross-Correlation (CC) and the Mutual Information (MI) between their dynamical
trajectories \cite{tsonis_2006,victor_prl_2007,tsonis_2008,yamasaki_2008,kurths_epl_2009,
bialonski_chaos_2010,lai_prl_2010,barreiro_chaos_2011,palus_chaos_2011,havlin_2012,
stanley_pre_2012}. Depending on the field of application, the choice of similarity estimators
is wider and includes partial correlations, graphical models, and adapted estimators, such
as event synchronization \cite{quiroga_2002} (recently used to analyze the summer monsoon
rainfall over the Indian peninsula \cite{kurths_2012}) or response dynamics \cite{timme_2007,
timme_2011}. However, any similarity measure used to compare two time-series usually results
in a non-zero value \cite{kurths_2002,panzeri_2007,bjorn_2010,geisel_2011,kurths_2013}. A
reason for this is that, in finite data sets, the presence of persistent trends and/or
deterministic recurrent oscillations results in spurious correlations \cite{palus_2007,
palus_2011,haam_2012}. Therefore, network reconstruction methods use fixed link densities
(where only the strongest similarity estimates are retained as links, e.g. in
\cite{marwan_2014,deza_2014,feng_2014}), link weights (where links are weighted based on
the similarity \cite{havlin_2011}) or pairwise significance tests \cite{kurths_2014} to
ensure the link representativeness. Another reason, which is the focus of our work, is the
network connectivity. In particular, the existence of teleconnections \cite{kurths_2014}
(name given in paleoecology to the long-range connections) in systems with multiple/continuous
coupling structures result in high similarity estimates between distant nodes. Moreover, even
after detrending a data set, the connectivity of the underlying network topology still plays
a major role in the non-zero values of the similarity measures between nodes if the network
is connected.

Indeed, in undirected connected complex networks, which are the focus of this work, every
pair of units is joined by some path. Consequently, any pair of units will exchange some
level of correlation or information due to the overall connecting topology. Therefore, the
ability to detect a \emph{direct} link between any two units is hindered within the
similarity measured value. Nevertheless, when the links are homogeneous, one expects that
directly connected units have larger values of the similarity measure than indirectly
connected ones \cite{proof_1}. Then, the existence of a certain threshold, $\tau$, that
can split the similarity values into two sets can be expected. If a similarity value between
two units is larger than $\tau$, it is considered to be significant and a consequence of a
direct link between the two units. Otherwise, it is less significant and a consequence of
the lack of a direct link between the two units. A similar method is used in paleoecology
to select good modern analog samples for climate and  environmental reconstruction
\cite{Duprat,Sawada}. When the strengths of the links are heterogeneous, the weak links are
further hindered within the similarity measure values and a bivariate analysis can be
insufficient \cite{bjorn_2009}. Moreover, in the presence of strong coupling, global patterns
in the system's behavior emerge, creating an effective topology which makes the underlying
network inference process unfeasible. Similarly, inference fails for very weak couplings,
where the system is hard to distinguish from being composed of uncoupled units. We find that
avoiding such fully coherent (large coupling strengths) or incoherent (small coupling
strengths) behavior is critical for the detection of the direct links (as it was also found
in Ref.~\cite{timme_2011}).

Since different topologies are inferred for different $\tau$ values, the problem of finding
the optimal $\tau$ value which recovers the largest portion of the underlying network is far
from trivial. For example, in Ref.~\cite{lai_prl_2010} the presence of dynamical noise in the
individual units was shown to enable the identification of an optimal threshold giving an
accurate prediction of a network topology, based solely on the measurement of dynamical
correlations. However, the method requires computing the inverse matrix of the dynamical
correlation matrix, which can be computationally demanding, and also the influence of
non-additive noise and/or observational noise remains an open question. Other methods for
link identification are based on perturbing the individual units. For example, the method
proposed in Ref.~\cite{pik_PRL_2011} requires performing independent, simultaneous, and
random phase resettings in all the units, which can be impractical in many real-world
systems (such as in \cite{marwan_2014,deza_2014,feng_2014,havlin_2011,kurths_2014,runge_2011}).

In this work, we show that, when the ordered values of CC (computed in absolute value, i.e.,
the Pearson coefficient) and/or MI (computed via ordinal pattern analysis \cite{Bandt2002,
Amigo,rosso_2012,st_2013}) exhibit a discontinuous curve, an adequate $\tau$ value permits
\emph{inference of underlying topologies without errors}. The exact link detection is
demonstrated by considering various discrete-time dynamical units (logistic maps, circle maps,
tent maps, and novel maps modelling laser arrays referred to as optical maps) that mutually
interact in different coupling topologies, including random networks. This means that, when
the discontinuity is observed, both methods are able to infer the exact underlying network
topology that interconnects the units from the local time-series measurements. As a result,
the topology of the interacting system is directly related to its function. We find that the
existence of this $\tau$ occurs even when observational noise and heterogeneities (in the
links and/or in the units) are present.

Our results are based on a critical comparison of the CC and MI inference methods
\emph{effectiveness} (what is the portion of the underlying topology that is reconstructed
correctly), \emph{robustness} (how the effectiveness is affected when parameters are changed,
namely, the heterogeneity in the map's dynamics or network weights, the coupling strength
between maps, and the network size or connectivity), and \emph{reliability} (results yield
consistent inferred networks, even when including observational noise and reducing the
time-series lengths). We conclude that MI outperforms CC as the most robust, in particular,
is the least sensitive to the choice of $\tau$ value, and reliable measure. To the best of
our knowledge, such reliable reconstruction of network topologies without errors from
time-series measurements of discrete-time dynamical units has not been previously obtained.

 \section{Model and Methods}
We consider logistic maps, tent maps, circle maps, and novel maps recently proposed
\cite{Roy_NatPhys_2012} for representing non-linear optical elements (in the following,
referred to as \emph{optical maps}). We observe that our method of network inference is
not restricted to maps, but it is demonstrated with maps mainly because of two reasons.
First, maps are computationally cost-efficient, allowing long time-series simulations,
performing robust statistical analysis, and have been widely used to study complex networks
of coupled units \cite{neiman_1999,Marti,Ponce2009}. Second, continuous systems can be
represented by maps (for example, there are many maps that represent various types of neurons
\cite{neiman_1999,ibarz_2011}) or transformed into maps by means of a Poincar\'e section or
a stroboscopic sampling (time-Poincar\'e). Here, we let the units have a degree of
heterogeneity by using non-identical parameters. For the underlying topology, we use random
networks (RN) \cite{Erdos1959} and small-world networks (SW) \cite{Strogatz1998} with
homogeneous and heterogeneous weights. These networks are characterized by the number of
nodes, $N$, the connectivity parameter, $p$, and the weights heterogeneity degree, $g$
(details are provided in the Supplemental Material \cite{SuppMat}).

 \subsection{Network topologies}
\begin{figure}[htbp]
 \begin{center}
  \textbf{(a)} \\
  \includegraphics[width=0.9\columnwidth]{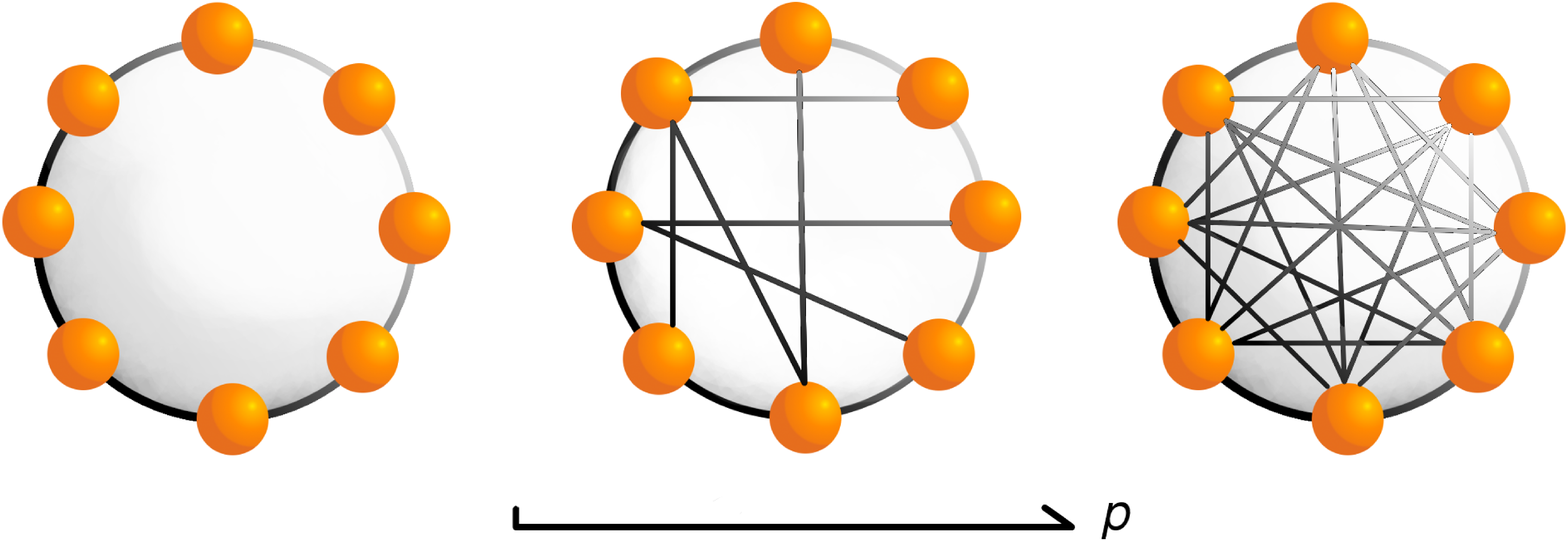} \\
  \textbf{(b)} \\
  \includegraphics[width=0.9\columnwidth]{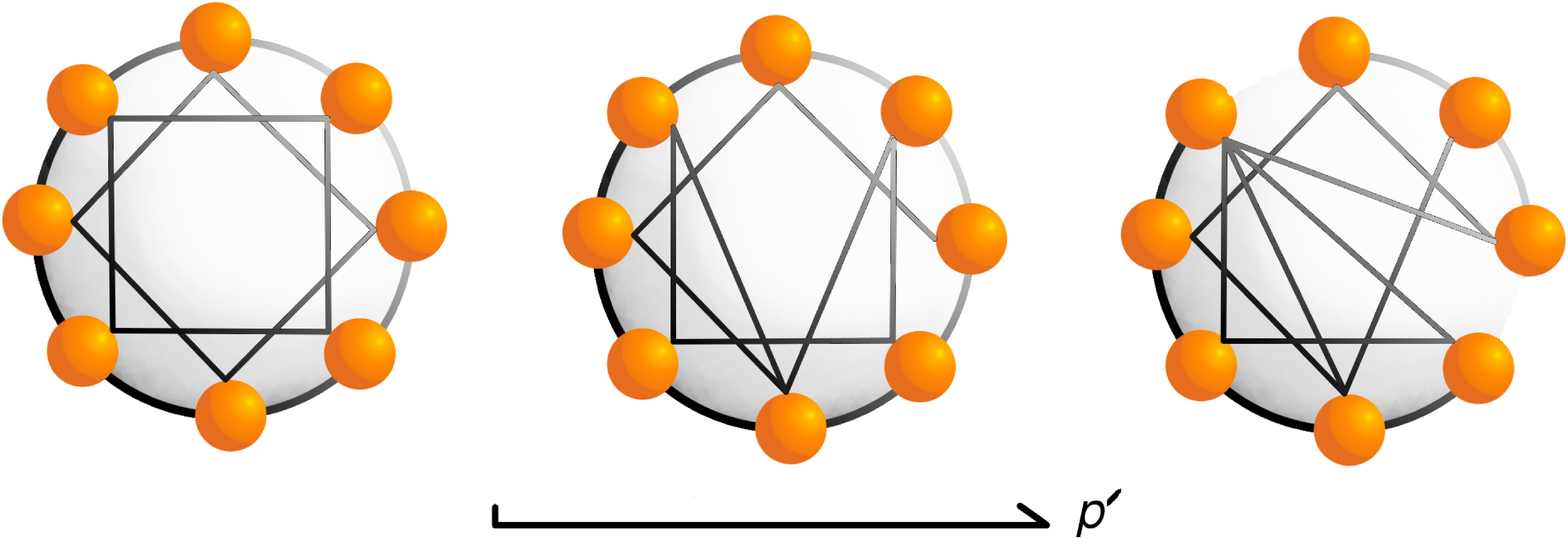}
 \end{center}
  \caption{Schematic diagrams for the generating processes of a random network [panel
  {\bf (a)}] and small-world network [panel {\bf (b)}]. The generation of the random
  (small-world) network is a function of the probability $p$ ($p'$) of adding (rewiring)
  links to a ring (regular) graph of $N$ nodes.}
 \label{fig_networks}
\end{figure}

Our Random Networks (RN) are characterized by the probability $p$ of adding a link between
every disjoint pair of nodes on a ring graph of $N$ nodes [Fig.~\ref{fig_networks}{\bf (a)}]
\cite{newRN}. For $p = 0$ we have a ring graph and for $p = 1$ an all-to-all network, while
random graphs are obtained for intermediate values of $p$ (where the Wigner semicircle
distribution of eigenvalues is achieved and the node degrees are Poisson distributed). On
the other hand, our Small-World networks (SW) are characterized by the probability $p'$ of
rewiring each link of a regular graph of degree $k = N/4$ [Fig.~\ref{fig_networks}{\bf 
(b)}], as in Ref.~\cite{Strogatz1998}.

\begin{figure}[htbp]
 \begin{center}
  \includegraphics[width=0.49\columnwidth]{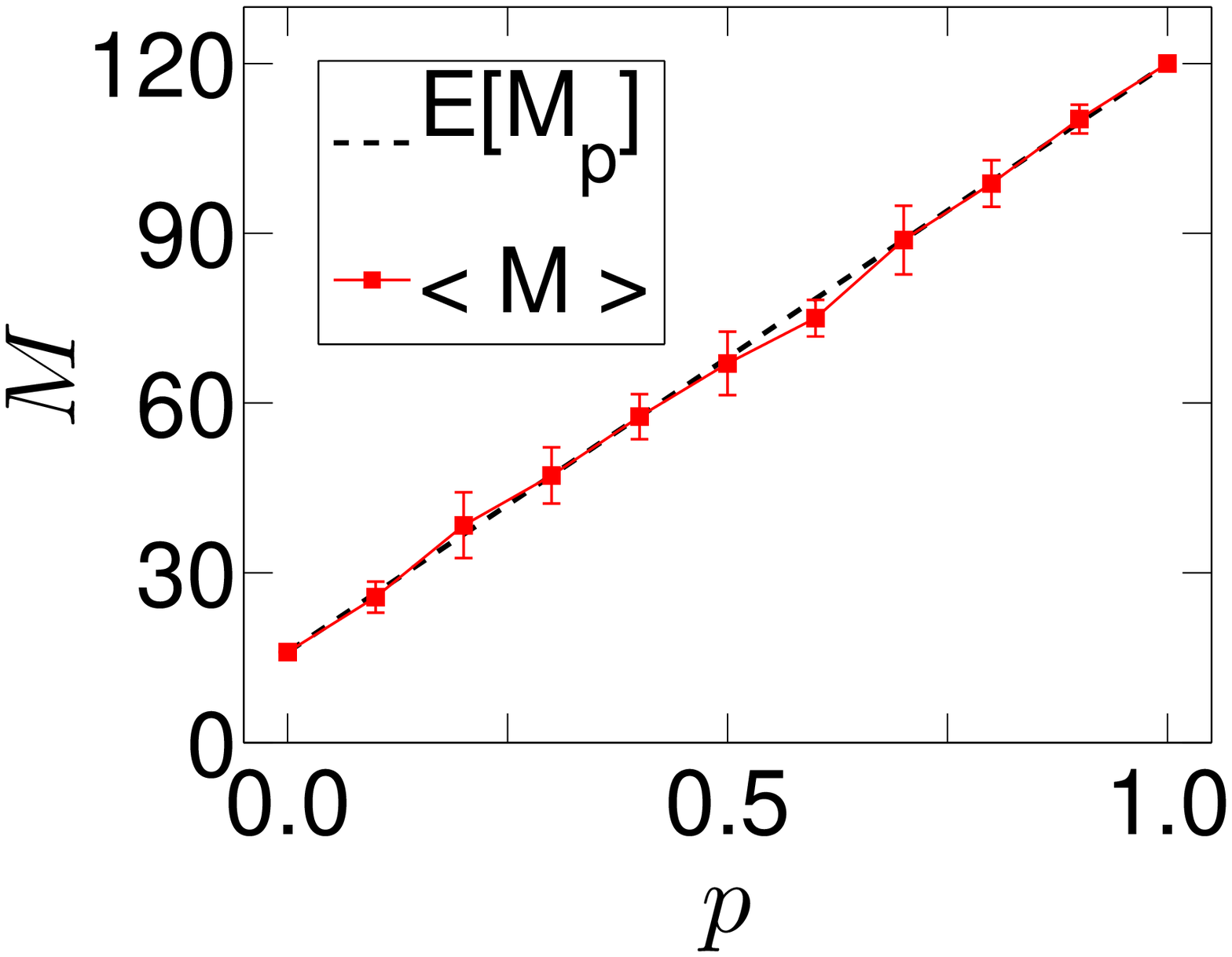}
  \includegraphics[width=0.49\columnwidth]{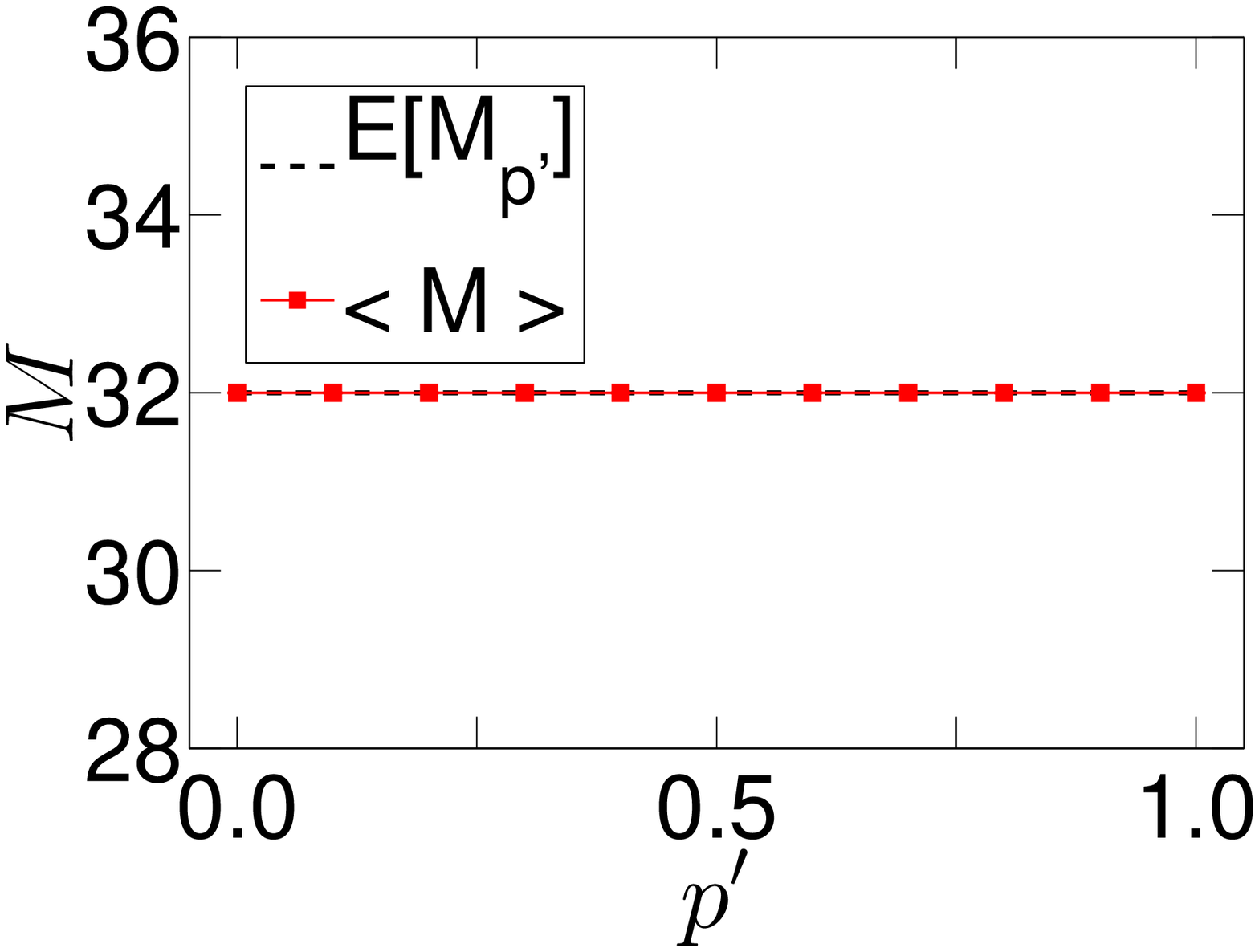}
 \end{center}
  \caption{Expected number of links ($E[M]$, dashed line) and average number of links
  ($\left\langle M \right\rangle$, square symbols) for $10$ of our network realizations
  as a function of the probability $p$. The error bars correspond to the standard deviation
  in the average number of links that the realizations have for each value of $p$. The left
  panel correspond to RNs of $N = 16$ nodes and the right panel corresponds to SW networks
  with $N = 16$ nodes.}
 \label{fig_RN_edges}
\end{figure}

These topologies define the underlying inter-connections of our network of coupled maps,
and our goal is to be able to infer them through the use of similarity measures. The choice
of these two types of networks is due to the difference in their number of links, denoted
by $M$. Our RN are mainly sparse networks for $p\sim0$ and $N$ large. For finite $p$, the
expected number of links is given by $E[M_p] = p\,N\,(N - 3)/2 + N$, which is the random
component plus the fixed ring structure (left panel in Fig.~\ref{fig_RN_edges}). Our SW
networks have a large number of links for large $N$, namely, $E[M_{p'}] = k\,N/2 = N^2/8$
for every $p'$ (right panel in Fig.~\ref{fig_RN_edges}). Consequently, the underlying number
of connections in our system changes appreciably when using RNs or SWs topologies.

The effect of coupling heterogeneity between the maps on the similarity measure inference is
dealt by using weighted networks. After an underlying topology is fixed, i.e., a particular
adjacency matrix $A_{ij}$ is set, random weights are associated to each existing direct link.
Specifically, we define a weighted network by
\begin{equation}
  W_{ij} = A_{ij}\,\left( 1 + g\,\xi_{ij} \right),\;\text{for}\,j>i\,,
 \label{eq_weights}
\end{equation}
where $1>g\geq0$ is the coupling heterogeneity degree parameter and $\xi_{ij}\in[-1,\,1]$ is an
uncorrelated zero-mean uniformly distributed random number. Symmetry in the weights is set
by using $W_{ij} = W_{ji}$, which keeps the links undirected.

  \subsection{Map's equation of motion}
The behavior of each map is governed by the equation
\begin{equation}
  x_{n+1}^{(i)} = \left( 1 - \epsilon \right)f\!\left(r_i,\,x_{n}^{(i)}\right) +
   \epsilon\sum_{j = 1}^N \frac{W_{ij}}{d_i}f\!\left(r_j,\,x_{n}^{(j)}\right),
 \label{eq_coupled_maps}
\end{equation}
where, $f(r,x) = r\,x\,(1 - x)$, for the logistic map, and $f(r,x) = x + r - 1.1\,\sin
\left(2\pi\,x \right)$ mod $1$, for the circle map (for other maps, see Supplemental Material
\cite{SuppMat}). $r_i$ is the $i$-th map parameter, $\epsilon$ is the coupling strength,
$W_{ij}$ accounts for the weight of the link [$W_{ij} = A_{ij}\,\left(1 + g\,\xi_{ij}\right)
= W_{ji}$, where $A_{ij} = A_{ji}$ is the adjacency matrix of the underlying topology,
$\xi_{ij}\in[-1,\,1]$ is an uncorrelated zero-mean uniformly distributed random number, and
$g$ is the degree of coupling heterogeneity], and $d_{i} = \sum_{j=1}^N W_{ij}$ is the
weighted degree of node $i$.

$N$ time-series are obtained from the trajectories of the $N$ maps, generated from
random initial conditions. Unless otherwise stated, the length of the time-series is $T =
5\times10^4$. The similarity measures are computed from these time-series. In particular,
the MI is computed from symbolic sequences of ordinal patterns of length $D = 4$
\cite{Bandt2002,Amigo,rosso_2012,st_2013} (see Supplemental Material for details
\cite{SuppMat}).

  \subsection{Similarity Measures and the threshold Method}
The zero-lag Pearson CC (referred only by CC on the following) is defined by
\begin{equation}
  CC_{ij} \equiv \frac{1}{T} \sum_{n = 0}^{T-1} \left[\frac{ x_{n}^{(i)} - \left\langle
   x^{(i)} \right\rangle }{\sigma_i} \right] \left[ \frac{x_{n}^{(j)} - \left\langle
    x^{(j)} \right\rangle }{\sigma_j} \right]\,,
 \label{eq_CC}
\end{equation}
where $\sigma_i$ is the $i$-th map time-series standard deviation, $\left\langle x^{(i)}
\right\rangle = \frac{1}{T} \sum_{n = 1}^T x_n^{(i)}$ is the time average of node's $i$
orbit, and $T$ is the number of iterations that the orbit has. In particular, \emph{we use
the absolute value of the CC} as the similarity measure for the inference process.

The MI is defined by transforming the time-series $\{ x_n^{(i)} \}_{n=0}^T$ into a symbolic
sequence and then calculating the probability of appearance of each symbol in the sequence.
The symbolic transformation we use is the ordinal analysis \cite{Bandt2002,Amigo}. The
\emph{ordinal analysis} transforms a length $D$ sliding window of each time-series, e.g.,
the vector $\{ x_n^{(i)},\ldots,\,x_{n+D-1}^{(i)} \}$, into a symbol $\alpha_n^{(i)}= 1,\,
\ldots,\,D!$. The symbol is the number of permutations needed to order the components of the
vector in a set of strictly increasing values. This means that each map's trajectory is
encoded into a sequence of $L \simeq T/D$ symbols if the sliding windows are non-overlapping
(which is the choice we make to have equally probable symbols in the case where the
time-series is random, e.g., for the surrogates of the maps trajectories).

MI is then calculated from
\begin{equation}
  MI_{ij} \equiv \sum_{\alpha_i,\,\beta_j = 1}^{D!} P\!\left( \alpha_i,\beta_j \right)
   \log_2\!\left[ \frac{ P\!\left( \alpha_i,\beta_j \right) }{ P\!\left(\alpha_i \right)
    P\!\left( \beta_j \right) }\right]\,,
 \label{eq_MI}
\end{equation}
where $P\left( \alpha_i \right)$ [$P\left( \beta_j \right)$] is the probability of having a
particular symbol $\alpha_i = 1,\ldots,D!$ [$\beta_j = 1,\ldots,D!$] in the encoded sequence
of map $i$ [$j$], namely, the frequency that $\alpha_i$ [$\beta_j$] appears in the encoded
$i$-th [$j$-th] map trajectory. Similarly, $P\left( \alpha_i,\,\beta_j \right)$ is the joint
probability that map $i$ has a symbol $\alpha_i$ and map $j$ a symbol $\beta_j$ (possibly
different than $\alpha_i$) in each encoded trajectory at equal times.

The choice of encoding is supported due to its simple implementation on experimental data
and robustness under noisy time-series observations. Furthermore, the encoding is implemented
without the need to define arbitrary partitions of the system's phase space. Also, the
symbolic sequence is easily interpreted. For example, a periodic orbit of period $P$ is
transformed to a symbolic sequence with a unique symbol $\alpha$ if an embedding dimension
$D = P$ is used. In other words, only one of the possible $D!$ symbols appears in the
symbolic sequence of a periodic orbit of period $P = D$. In such a case, the symbolic
entropy of the sequence is zero ($H = - \sum_{\alpha = 1}^{D!} P(\alpha)\,\log_2[ P(\alpha)
] $), hence, the MI is also zero. When the period of the orbit is different than $D$, the
symbolic sequence has a non-null entropy, however, the joint entropy between two periodic
orbits is the sum of the entropies for each orbit (independent orbits). Consequently, $MI
= 0$ between two periodic orbits.

On the other hand, the CC value between periodic orbits depends on the phase-lag ($\phi$)
value between the two orbits. It is safe to say that for small phase-lags ($\phi \ll 1$)
$CC\sim1$ and for large phase-lags ($\phi\sim\pi$) $CC\sim-1$, hence, also close to unity
in absolute value.

The \emph{threshold}, $\tau$, used to split the similarity values, is a control parameter
that allows to define the \emph{inferred} adjacency matrix, $A_\tau$: $A_{\tau,\,ij} = 1$
if the similarity measure between maps $i$ and $j$ is larger than $\tau$, and $A_{\tau,\,
ij} = 0$ otherwise. The \emph{error}, $\Delta$, between the inferred and the true
adjacency matrices, is defined by
\begin{equation}
  \Delta = \frac{ \sum_{i,\,j=1}^N \left| A_{ij} - A_{\tau,ij} \right| }{N\,(N - 1)}\,.
 \label{eq_error_adj}
\end{equation}
The minimum value of $\Delta$ is $0$, which corresponds to an exact detection of the true
underlying topology. The maximum value of $\Delta$ is $1$, and occurs only if all the links
are inferred incorrectly.

The \emph{effectiveness} of a similarity measure (CC or MI) is quantified by the error
$\Delta$ between the true topology and the inferred topology, which is a function of the
particular threshold $\tau$ chosen. We also consider the receiver operating characteristic
(ROC) curve, which quantifies the true positive rate (TPR) and false positive rate (FPR),
each measure being a function of $\tau$ \cite{ROC_def,kurths_2013}. We consider that a
measure is effective when $\Delta \simeq 0$, the TPR is maximum, and the FPR is minimum.
The \emph{robustness} of the CC or MI is quantified in terms of how sensitive $\Delta$ is
to changes in the system's parameters (map's parameter, network topology, and heterogeneity
degree) and choice of $\tau$ value. We also analyze how $\Delta$ depends on the length of
the time-series, the level of observational noise, and the size of the network. We consider
that a measure is robust, when small changes to the system's parameters or optimal $\tau$
value keep $\Delta\simeq0$. The \emph{reliability} of a method is the ability to give
consistently similar results from similar observations; hence, a measure's reliability gives
an estimation of the reproducibility of the results.

 \section{Results}
In the following we present results for chaotic logistic maps ($r = 4$) and circle maps ($r
= 0.35$) coupled in RNs with $N = 16$ and $g = 0.1$. Results for other maps, coupled with
other network topologies and heterogeneity degrees, are found in the Supplemental Material
\cite{SuppMat}.

\begin{figure}[htb!]
 \centering
  \begin{minipage}{9pc}
   \small{\textbf{(a)}}\\
   \includegraphics[width=9pc]{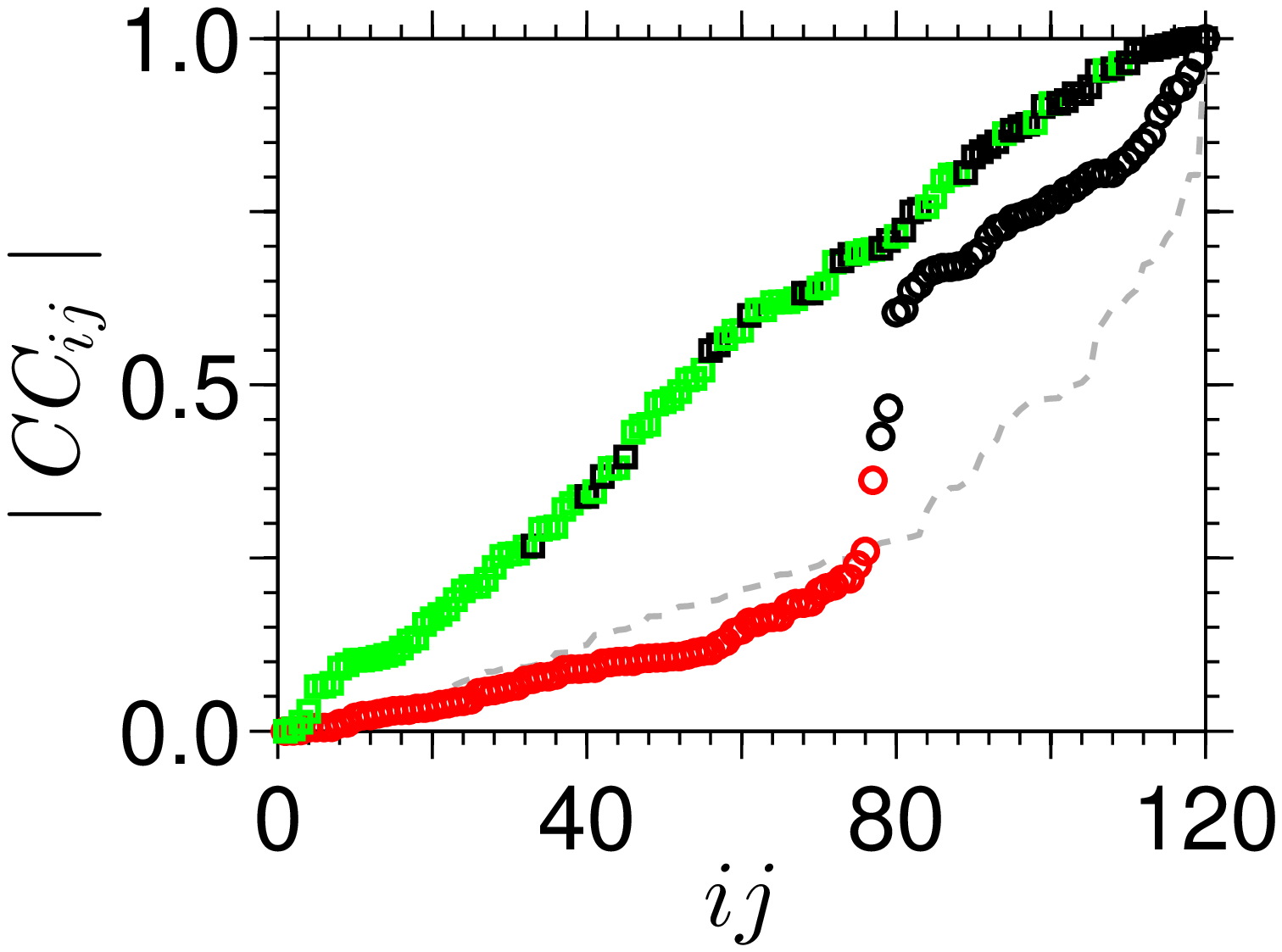}
  \end{minipage} \hspace{1pc}
  \begin{minipage}{9pc}
   \small{\textbf{(b)}}\\
   \includegraphics[width=9pc]{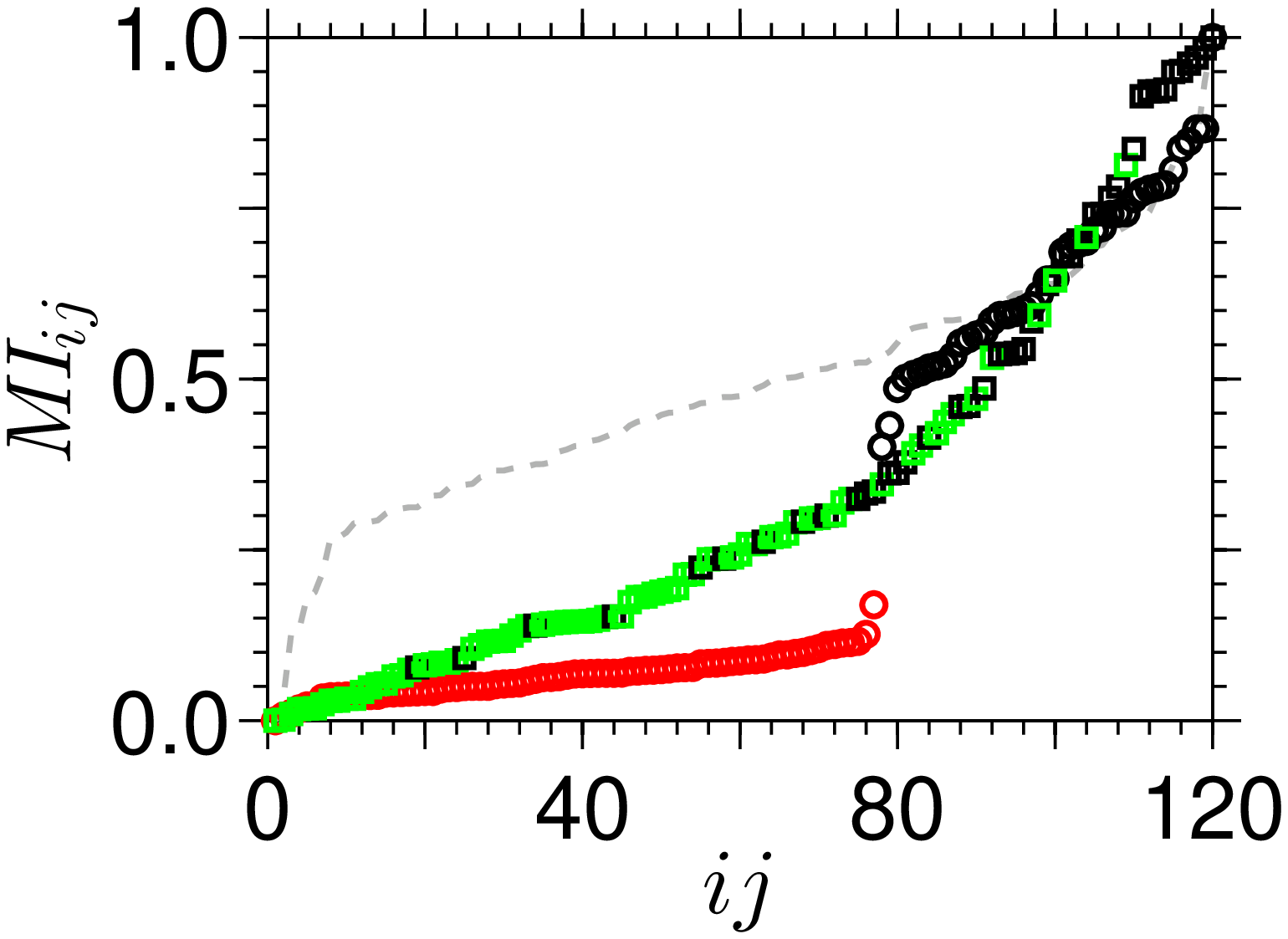}
  \end{minipage}
 \caption{(Color online) Panel {\bf (a)} [Panel {\bf (b)}] shows the normalized CC [MI]
 values, ordered from smaller to larger values, for all pair of nodes in a weighted ($g
 = 0.1$) random network ($p = 0.3$) of $N = 16$ identical chaotic ($r_i = 4.0\;\forall\,i$)
 logistic maps. The dashed curves (gray online) correspond to the uncoupled situation
 ($\epsilon = 0$), where no discontinuity is observed. The circles (squares) correspond to
 a coupling strength of $\epsilon = 0.06$ ($\epsilon = 0.5$), where the values for direct
 links are signaled by dark --black online-- circles (squares) and the values for indirect
 links are signaled by light --red (green) online-- circles (squares). The discontinuity is
 again absent for $\epsilon = 0.5$ due to the coherent dynamical behavior (synchronous
 orbits) that the system exhibits at this stage.}
 \label{fig_1}
\end{figure}

Figure~\ref{fig_1} shows the ordered values of the normalized CC [Fig.~\ref{fig_1}{\bf (a)}]
and MI [Fig.~\ref{fig_1}{\bf (b)}] measures for a particular RN ($p = 0.3$) of identical
($\delta r = 0$) logistic maps coupled with $\epsilon = 0.06$ (circles) and $\epsilon =
0.5$ (squares). Discontinuous curves are found for both, CC and MI, for $\epsilon = 0.06$,
though, the gap for MI is larger than the one for CC. We observe that in this case the direct
connections (indicated by darker --black online-- symbols) are found to have large similarity
values, while the indirect connections (indicated by lighter --green online-- symbols) have
lower values. For comparison, the CC/MI values for $\epsilon = 0$ are shown in light --gray
online-- dashed lines.

\emph{As a general result, we note that the effectiveness of a similarity measure to infer
the underlying topology relies on the existence of a discontinuous jump in its ordered
values.} The abrupt change corresponds to a difference between the values of the similarity
measure for direct connections and the values for indirect connections. Specifically, we
find that if a gap in the ordered sequence of CC or MI values exists, any value of $\tau$
within this gap infers the underlying topology without errors. When the gap is missing (as
for $\epsilon = 0.5$), the direct and indirect connection values are mixed within a
continuous curve (represented by squares in Fig.~\ref{fig_1}), hence, the error $\Delta > 0$
for any $\tau$ value. \emph{We find that a necessary condition for the appearance of the gap
is to avoid the fully coherent} (large coupling strengths) \emph{or incoherent} (small
coupling strengths) \emph{behaviors}.

\begin{figure}[htb!]
 \centering
  \begin{minipage}{9pc}
   \small{\textbf{(a)}}\\
   \includegraphics[width=9pc]{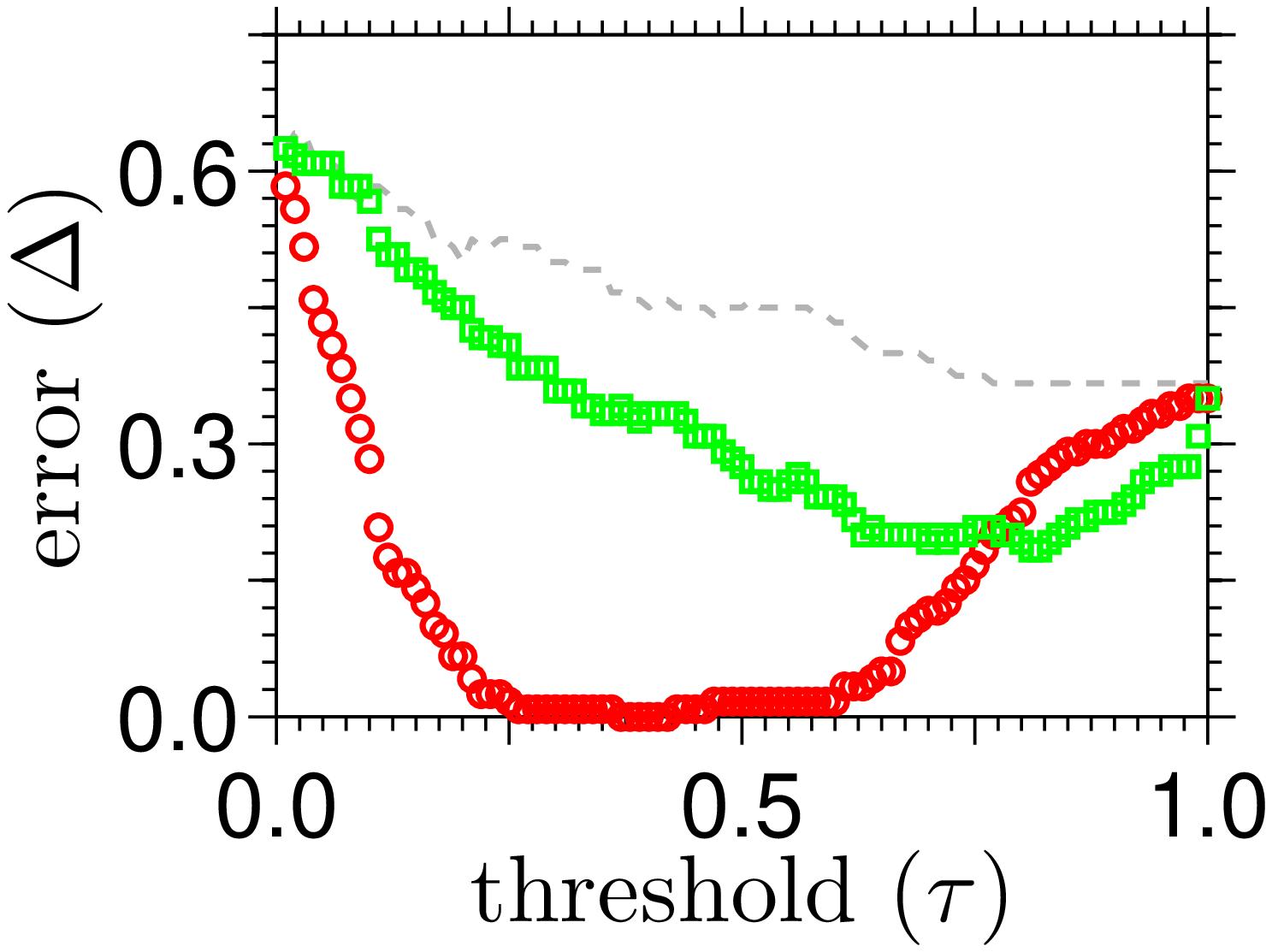} \\
   \small{\textbf{(c)}}\\
   \includegraphics[width=9pc]{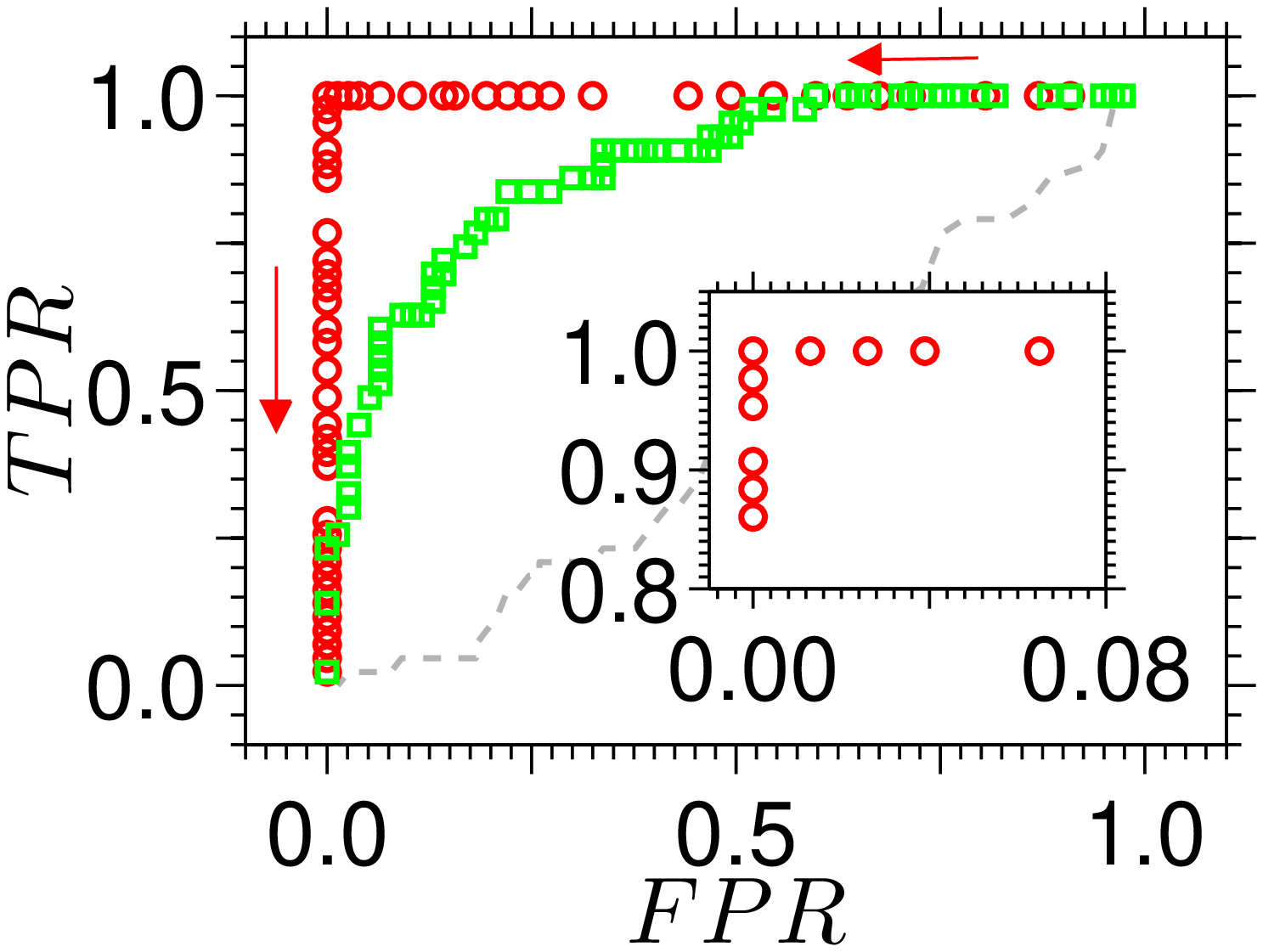} \\
  \end{minipage} \hspace{1pc}
  \begin{minipage}{9pc}
   \small{\textbf{(b)}}\\
   \includegraphics[width=9pc]{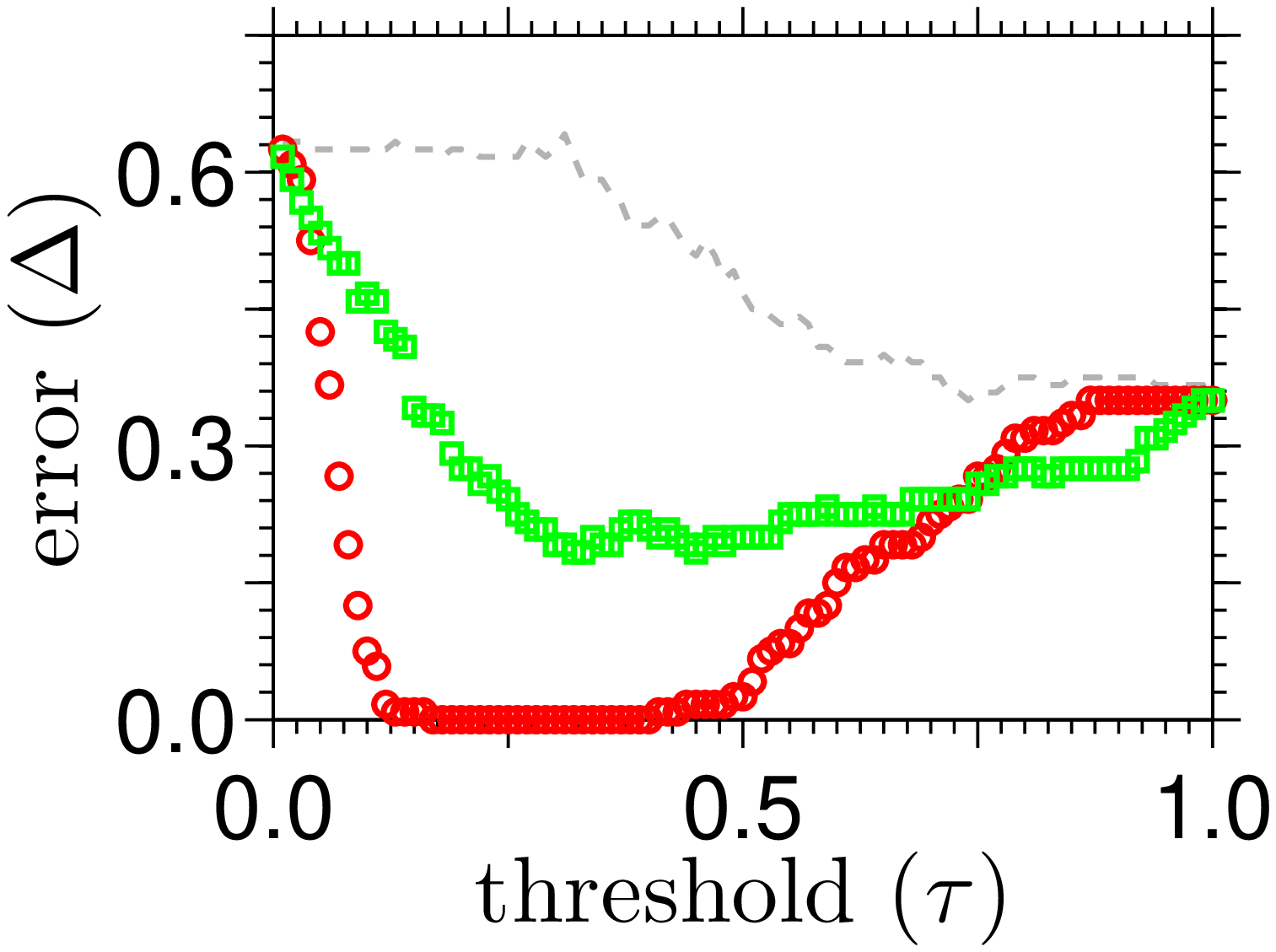} \\
   \small{\textbf{(d)}}\\
   \includegraphics[width=9pc]{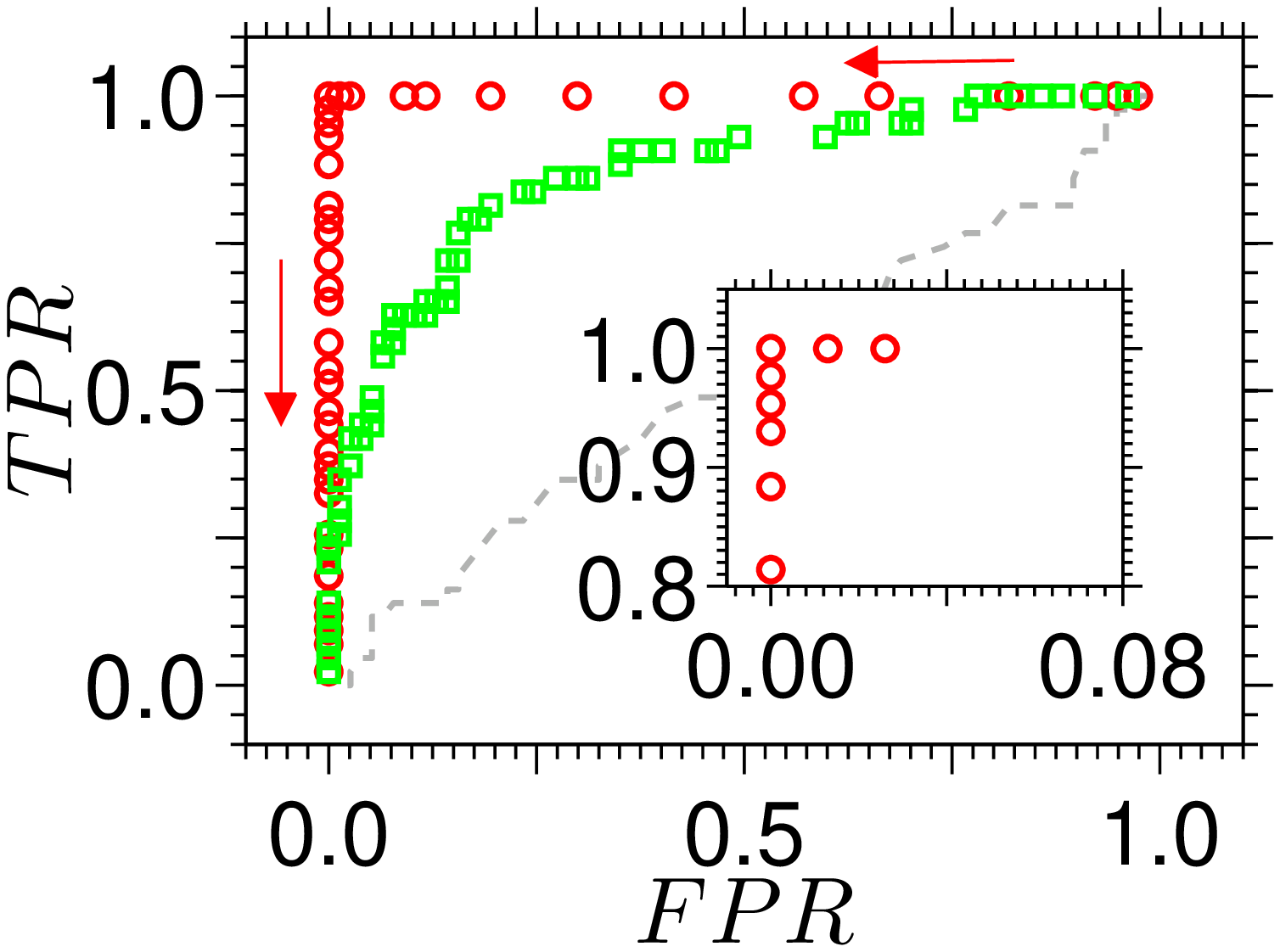} \\
  \end{minipage}
 \caption{(Color online) Panel {\bf (a)} [Panel {\bf (b)}] shows the CC [MI] inference
  error, $\Delta$, as a function of the different threshold, $\tau$, values for the
  curves and systems of Fig.~\ref{fig_1} (light --gray online-- dashed line for $\epsilon
  = 0$, dark --red online-- circles for $\epsilon = 0.06$, and dark --green online-- squares
  for $\epsilon = 0.5$). The corresponding ROC curves are shown in panel {\bf (c)} [panel
  {\bf (d)}], where the arrows indicate the direction in which $\tau$ increases. The insets
  in these panels are a zoomed view of the upper left corner of the ROC diagram, showing
  that for $\epsilon = 0.06$, both similarity measures achieve a perfect inference.}
 \label{fig_2}
\end{figure}

In Fig.~\ref{fig_2}{\bf (a)} and {\bf (b)} we indeed observe that the optimal choice of
$\tau$, namely, when $\Delta = 0$, is achieved for values of $\tau$ falling within the
discontinuity gap of Fig.~\ref{fig_1}{\bf (a)} and {\bf (b)}, respectively. Hence,
Fig.~\ref{fig_2} shows the effectiveness and $\tau$-robustness of the two inference
measures (in normalized units) for the same network of coupled maps as in Fig.~\ref{fig_1}.
The underlying network ($N = 16$ node RN with $p = 0.3$) has $M = 50$ links, out of a total
of $M_t = N\,(N-1)/2 = 120$ possible links. For $\tau = 0$, the inferred network has a
global all-to-all coupling topology (as all normalized CC/MI values are $\geq0$), therefore,
the error is the relative number of extra links detected (the false positives), $\Delta_{
\tau = 0} = ( M_t - M )/M_t = 70/120 \simeq 0.58$. For $\tau = 1$, the error is the number
of true links missed (the false negatives), and these are all the links, because the
inferred topology is a fully disconnected graph (as all normalized CC/MI values are $\leq1$),
therefore, $\Delta_{\tau = 1} = M/M_t= 50/120 \simeq 0.42$.

\begin{figure}[htb!]
 \centering
  \begin{minipage}{9pc}
   \small{\textbf{(a)}}\\
   \includegraphics[width=9pc]{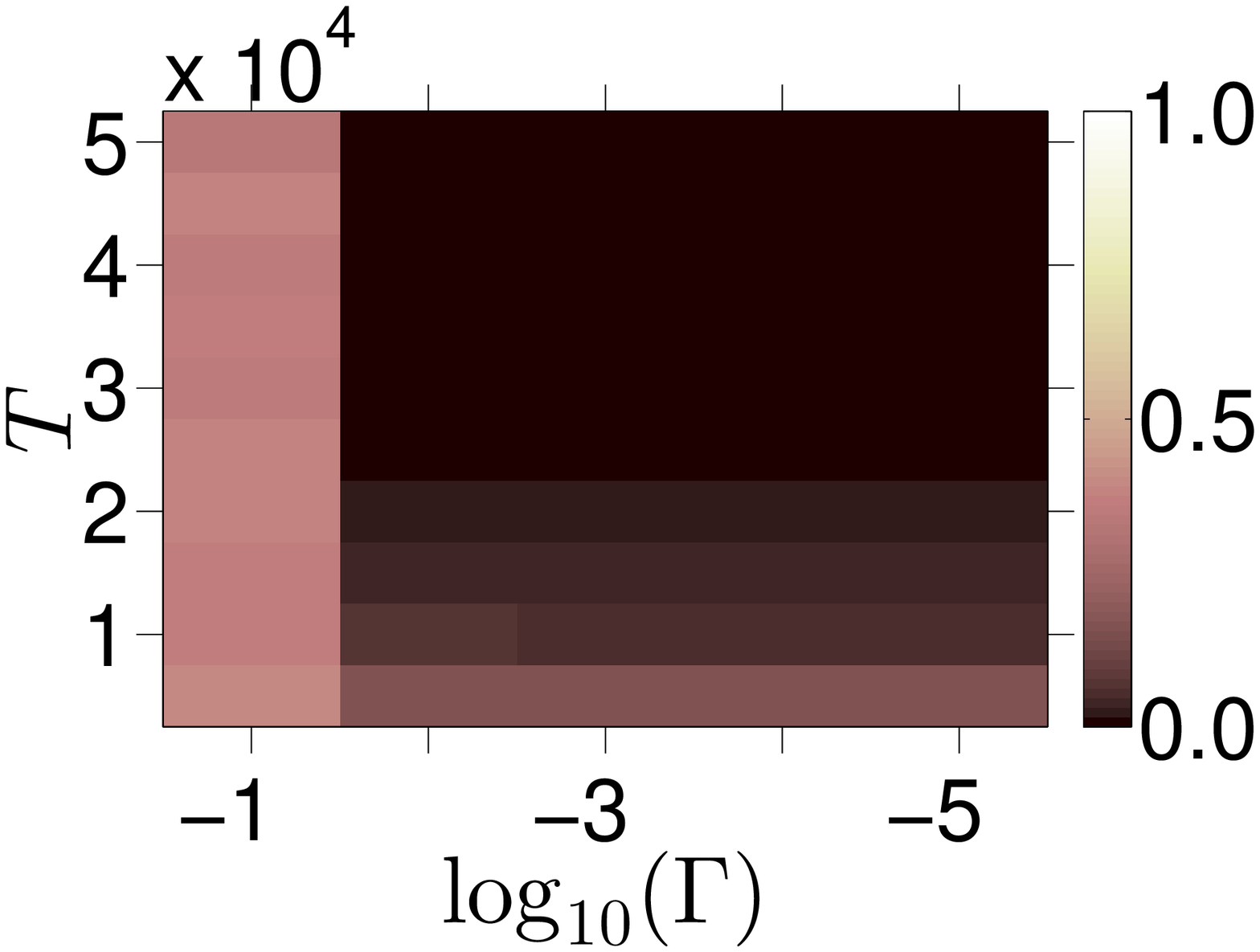}\\
   \small{\textbf{(c)}}\\
   \includegraphics[width=9pc]{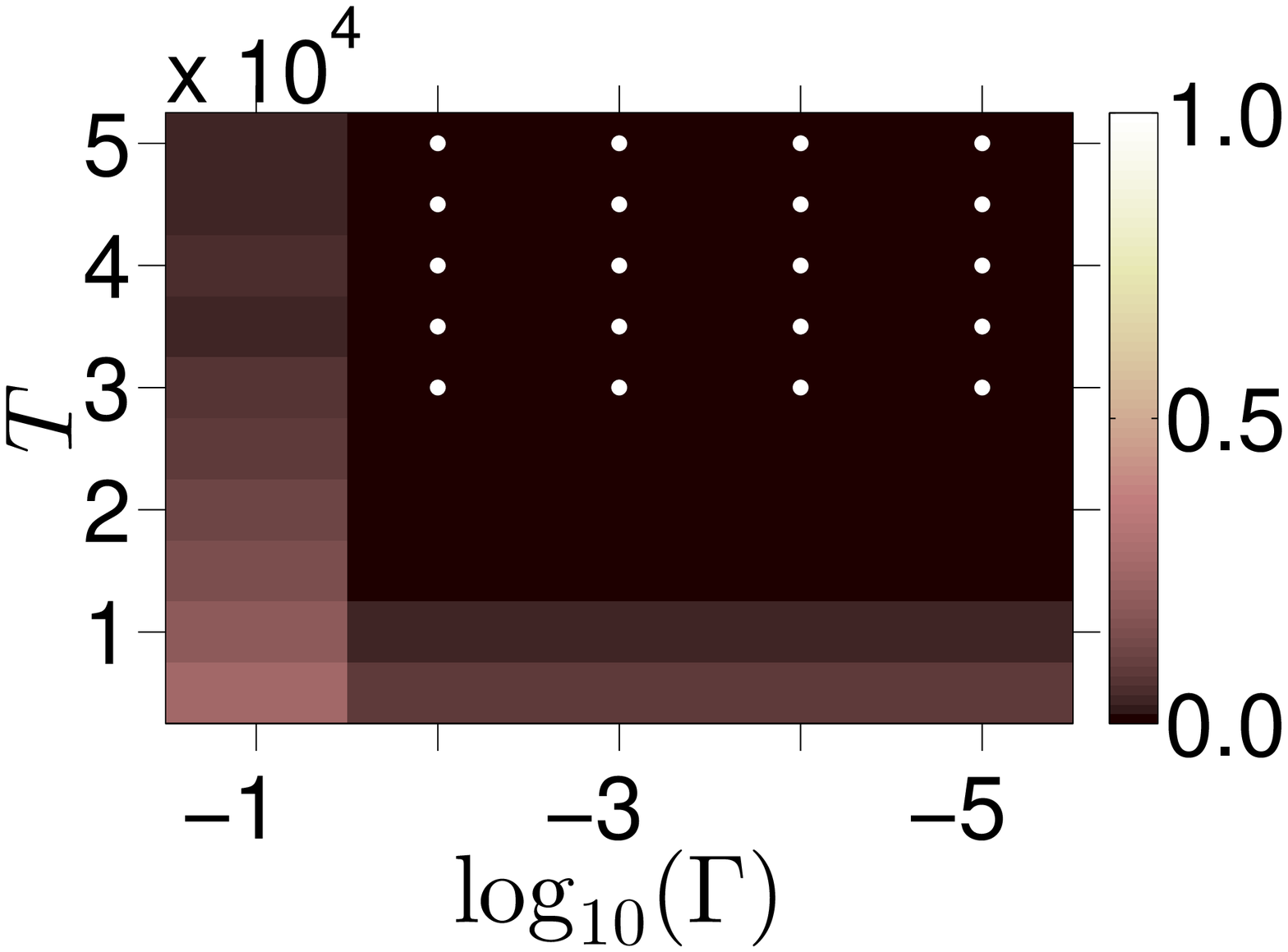}
  \end{minipage} \hspace{1pc}
  \begin{minipage}{9pc}
   \small{\textbf{(b)}}\\
   \includegraphics[width=9pc]{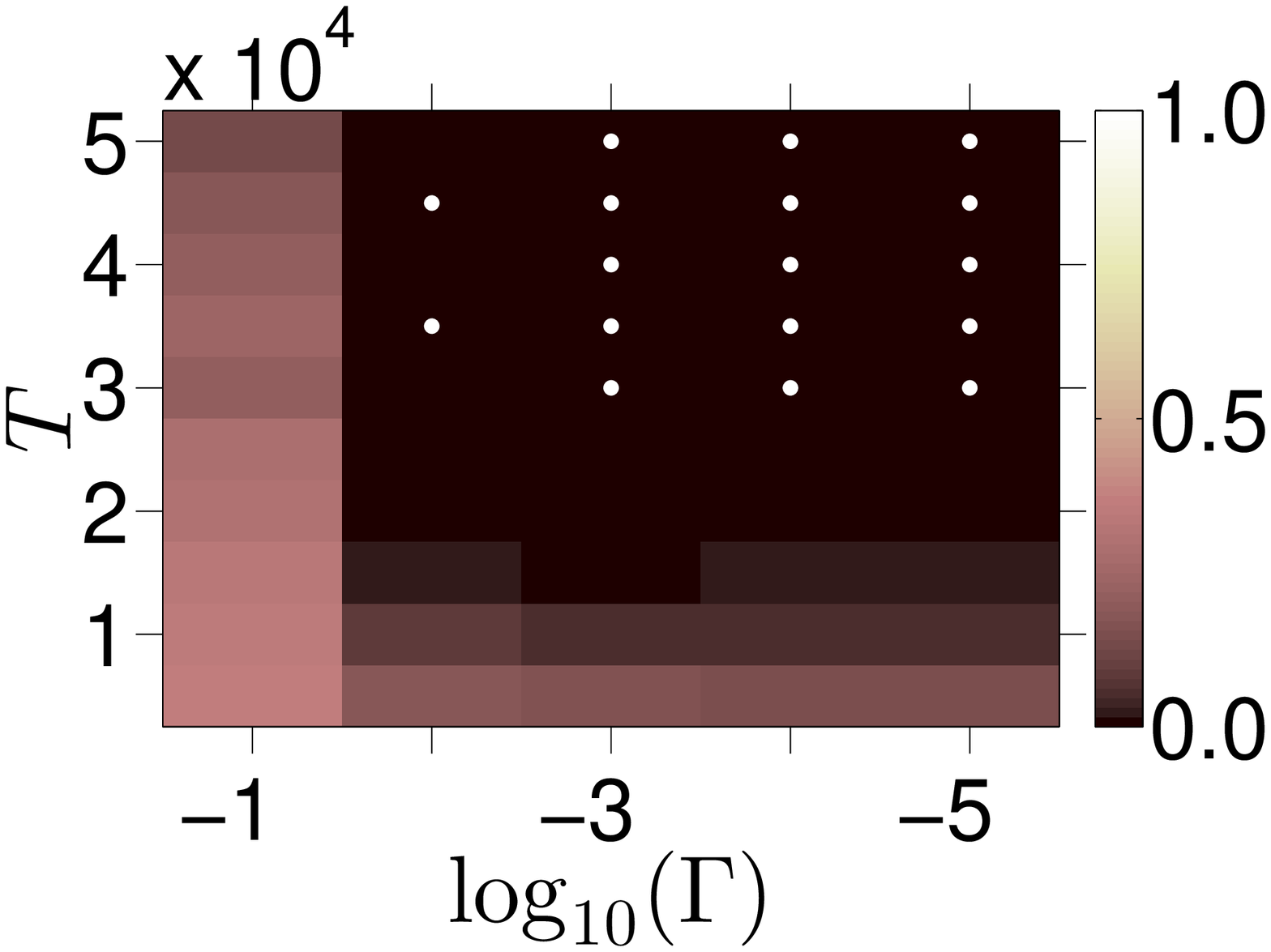}\\
   \small{\textbf{(d)}}\\
   \includegraphics[width=9pc]{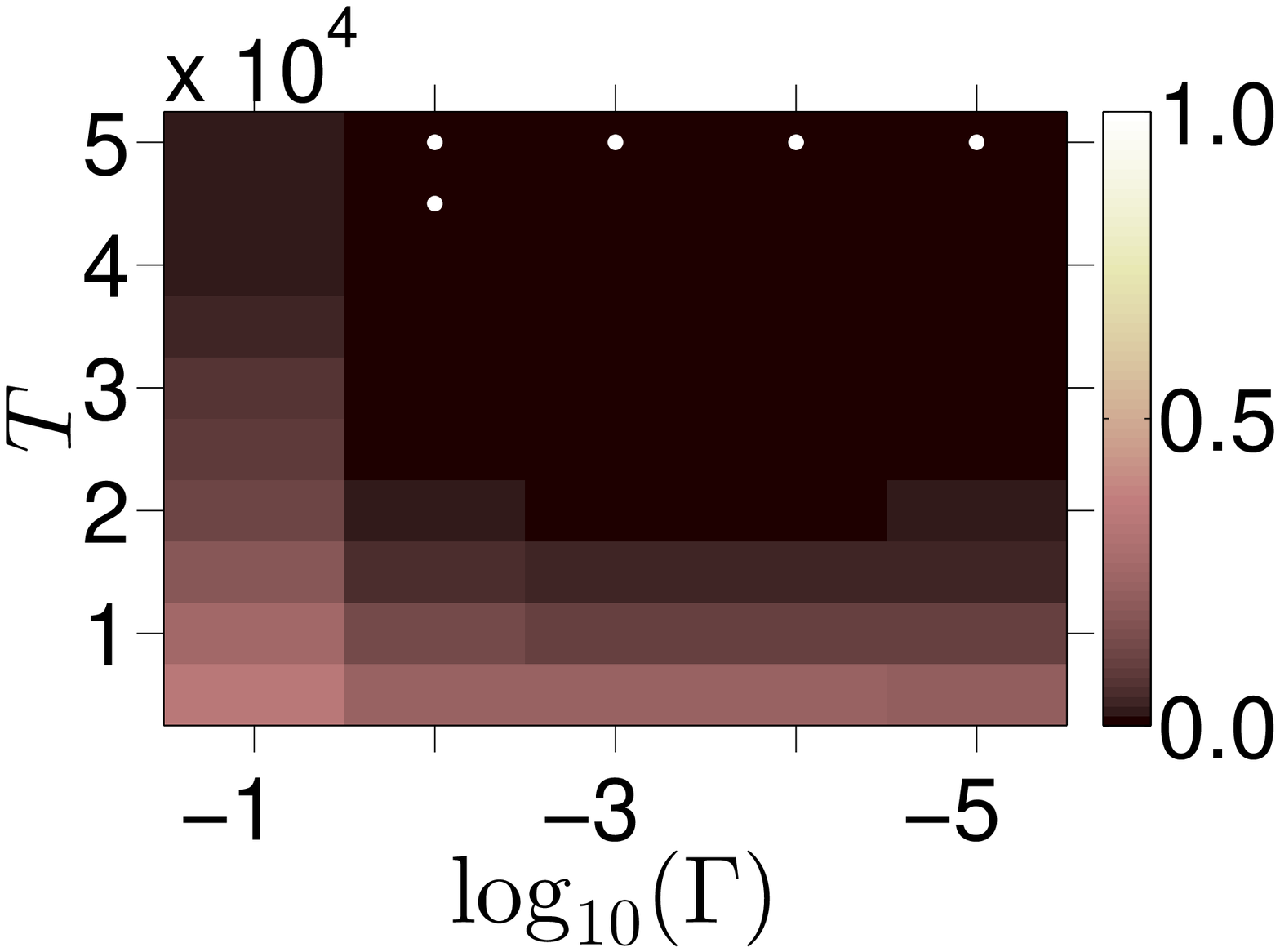}
  \end{minipage}
 \caption{(Color online) Minimal error [$\min(\Delta)$, in color code] values that are
  obtained from CC [panels {\bf (a)} and {\bf (c)}] and MI [panels {\bf (b)} and {\bf (d)}]
  measures of identically ($\delta r = 0$) chaotic logistic maps (top row) and circle maps
  (bottom row), averaged over $5$ RN realizations (each with equal network characteristics:
  $p = 0.3$, $N = 16$, and $g = 0.1$), as a function of the observational noise intensity
  ($\Gamma$) added to the times-series and its length ($T$). We set $\epsilon = 0.06$ for
  the logistic maps (as in Figs.~\ref{fig_1} and \ref{fig_2}) and $\epsilon = 0.12$ for the
  circle maps. The white dots indicate where an exact detection of all the underlying links
  is possible in all network realizations, i.e., $\min(\Delta) = 0$.}
 \label{fig_3}
\end{figure}

Most importantly, we found that, in general, MI is more robust than CC, in the sense that
it is able to recover the underlying topology without errors for more threshold values.
This is seen by comparing the lengths of the intervals where $\Delta = 0$ for $\epsilon =
0.06$ between Fig.~\ref{fig_2}{\bf (a)} for CC and Fig.~\ref{fig_2}{\bf (b)} for MI. The
wider interval is explained by the existence of a larger gap in the values that the MI curve
of circles has in Fig.~\ref{fig_1}{\bf (b)}, in contrast to the smaller gap in
Fig.~\ref{fig_1}{\bf (a)} [however, for some parameters, CC can be more robust than MI in
other aspects, as will be shown in Fig.~\ref{fig_3}{\bf (c)}].

Next, we consider the receiver operating characteristic (ROC) curves, which quantify the
true positive rate (TPR) and false positive rate (FPR) that each measure has as a function
of $\tau$ \cite{kurths_2013,ROC_def}. The ROC curves, shown in Fig.~\ref{fig_2}{\bf (c)}
and {\bf (d)}, provide further information about the type of errors made as a function of
the threshold. When $\epsilon = 0$, as the threshold increases from $\tau = 0$ to $\tau =
1$, both the TPR and FPR decrease (links are not inferred, regardless if they exist or not).
On the contrary, when $\epsilon > 0$, as $\tau$ increases, only the FPR decreases, while
all the existing links are correctly inferred (the TPR remains constant). When $\tau$ is
increased above the optimal range of values, then the TPR starts to decrease, as existing
links are not inferred (the FPR remains constant).

\emph{In situations where the knowledge of the underlying topology is missing}, the error
$\Delta$ [Eq.~(\ref{eq_error_adj})], TPR, and FPR, cannot be computed. However, if the
ordered values of the CC or MI exhibit a discontinuity gap, as in Fig.~\ref{fig_1}, the
links can still be divided into two sets. The links at the left of the discontinuity
(lower than the threshold value) correspond to indirect connections, while any value at
the right reveals a direct link (higher than the threshold value) \cite{proof_1}.
Every value of CC or MI inside the gap can be chosen as a possible threshold value $\tau$
and the width of the gap determines the sensitivity of the method. If the gap in the ordered
CC or MI values is absent, the method is not capable to infer the correct topology.

\emph{For practical applications in real-word data}, it is important to analyze how the
measures depend on the length of the data set, and on how noisy the data is. In the
following, observational noise is considered by adding uncorrelated zero-mean uniformly
distributed noise $\eta^{(i)} \in[-1,\,1]$ of strength $\Gamma$ to each data set.

Figure~\ref{fig_3} displays (in color code) how the \emph{minimal error} [i.e., $\min(
\Delta)$, the minimum value of $\Delta$ corresponding to an optimal $\tau$] depends on the
level of observational noise ($\Gamma$) and data availability ($T$) for the network
parameters of Figs.~\ref{fig_1} and \ref{fig_2} for identical ($\delta r = 0$) logistic
maps [Fig.~\ref{fig_3}{\bf (a)} and {\bf (b)}] and circle maps [Fig.~\ref{fig_3}{\bf (c)}
and {\bf (d)}]. Furthermore, to make the results reliable, we average the $\min(\Delta)$
value that is found for each $\Gamma$ and $T$ among $5$ RN realizations with equal statistical
characteristics. For the sake of clarity, the white dots in the darker regions indicate
where $\min(\Delta) = 0$, i.e., the perfect reconstruction of all the RNs.

We can see that MI infers exactly all the network realizations for moderate noise strengths
($\Gamma < 0.1$) and orbits with $T\geq3\times10^4$ for the logistic maps [white dots in
Fig.~\ref{fig_3}{\bf (b)}] and $T\geq5\times10^4$ for the circle maps [Fig.~\ref{fig_3}{\bf
(d)}]. However, we find that CC fails to provide reliable results for the logistic
maps, as it only infers the underlying topology for some RN realizations [dark color in
Fig.~\ref{fig_3}{\bf (a)}]. On the other hand, CC outperforms MI for circle maps [white
dots in Fig.~\ref{fig_3}{\bf (c)}]. In general, \emph{we note that both methods are
effective for moderate $\Gamma$ and $g$ when sufficient data is available}.

\begin{figure}[htb!]
 \centering
  \begin{minipage}{9pc}
   \small{\textbf{(a)}}\\
   \includegraphics[width=9pc]{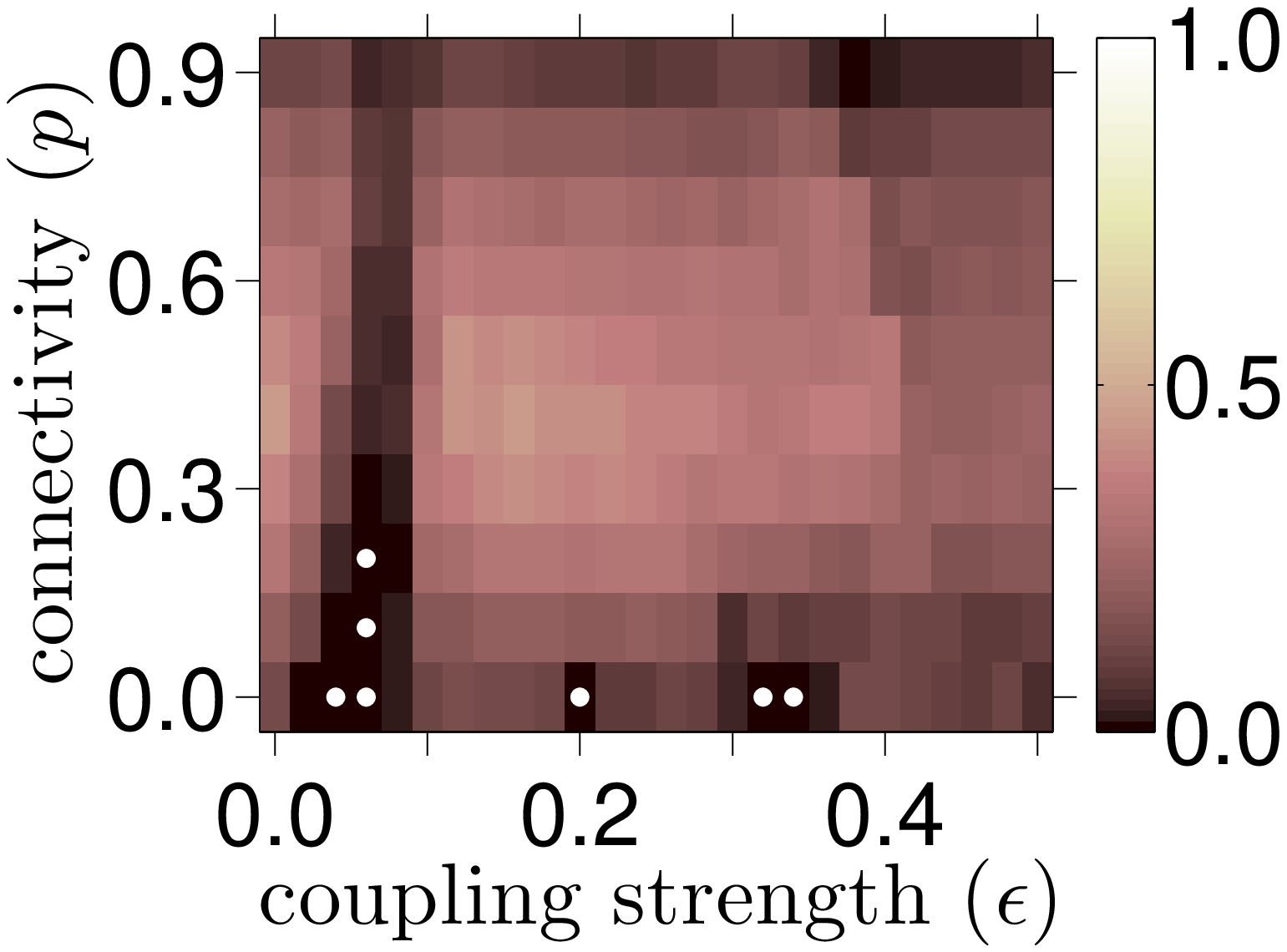}\\
   \small{\textbf{(c)}}\\
   \includegraphics[width=9pc]{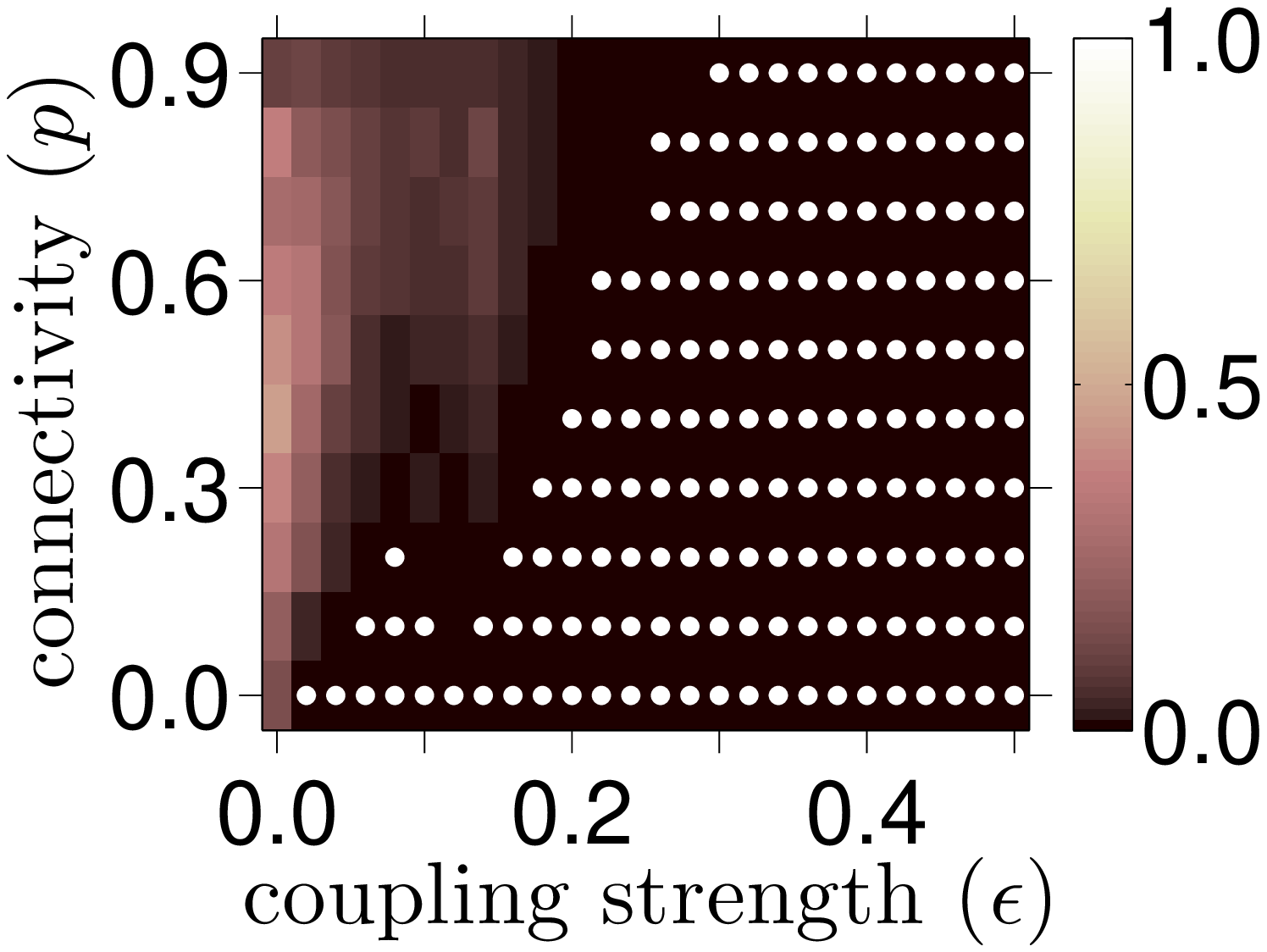}
  \end{minipage} \hspace{1pc}
  \begin{minipage}{9pc}
   \small{\textbf{(b)}}\\
   \includegraphics[width=9pc]{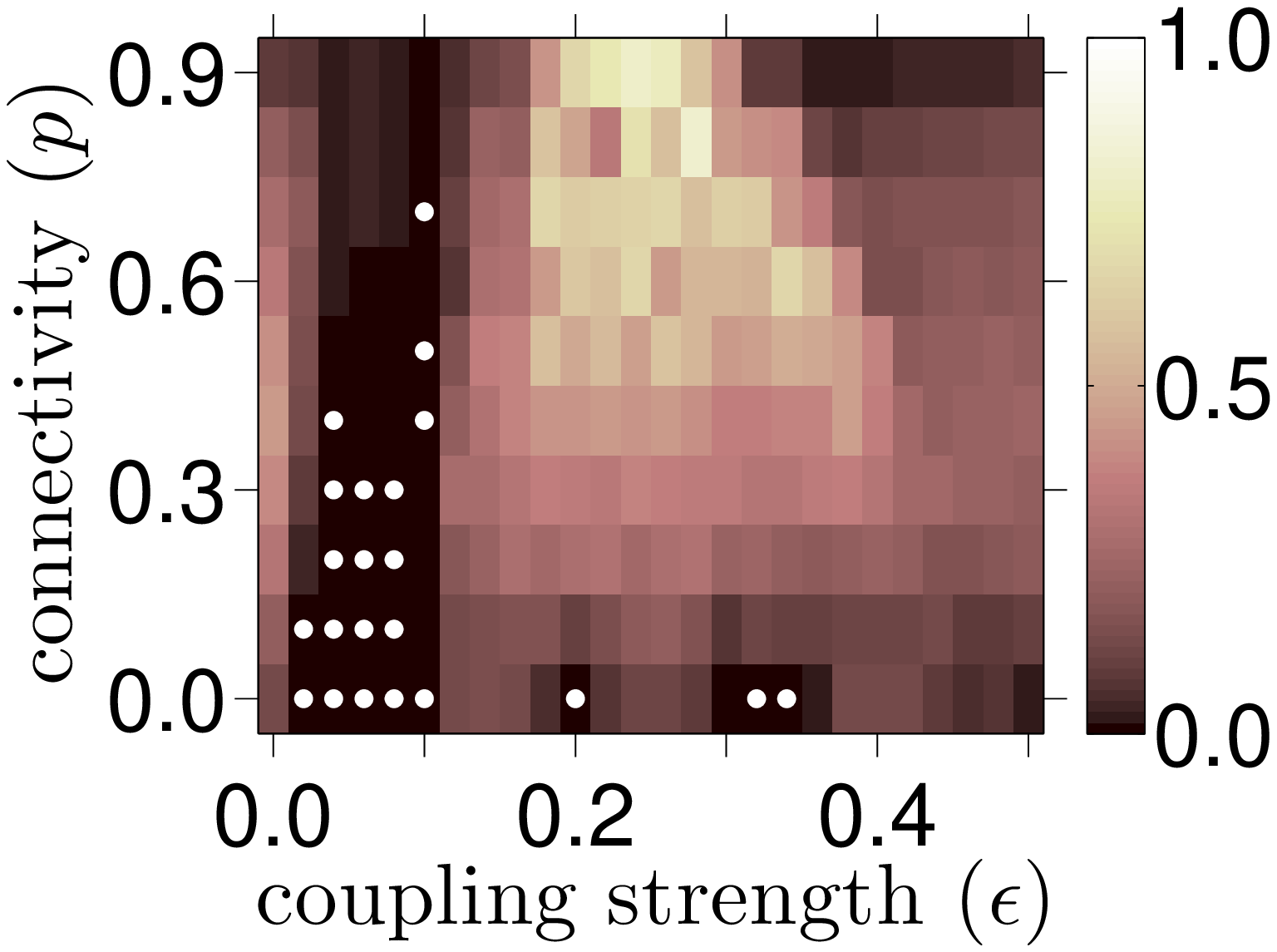}\\
   \small{\textbf{(d)}}\\
   \includegraphics[width=9pc]{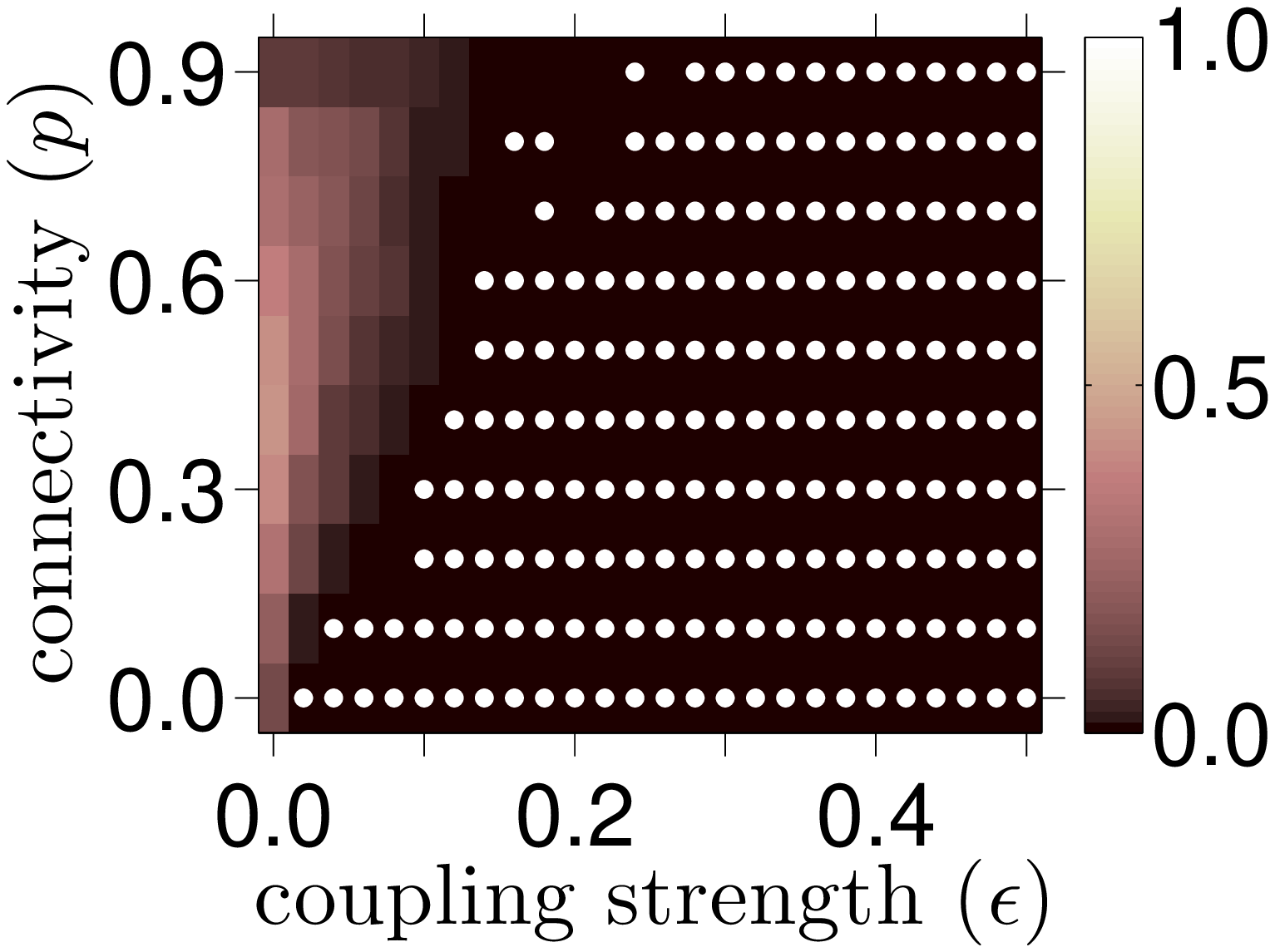}
  \end{minipage}
 \caption{(Color online) $\min(\Delta)$ values averaged (color code) over $5$ unweighed
  ($g = 0$) RN realizations with equal characteristics obtained from CC [panels {\bf (a)}
  and {\bf (c)}] and MI [panels {\bf (b)} and {\bf (d)}] measures for $N = 16$ non-identical
  ($\delta r = 0.1$) chaotic logistic (top row) and circle (bottom row) maps as a function
  of the network's connectivity parameter $p$ and coupling strength $\epsilon$. White dots
  indicate where $\min(\Delta) = 0$ in all networks.}
 \label{fig_4}
\end{figure}

Next we show that the exact detection of direct links is also possible when the individual
units are heterogeneous ($\delta r = 0.1$), for a wide range of RN parameters and coupling
strengths. This can be seen in Fig.~\ref{fig_4}, that displays the $\min(\Delta)$ (in color
code) as a function of the connectivity ($p$) and the coupling strength ($\epsilon$). Each
value of $\min(\Delta)$ is averaged over $5$ network realizations. Results for such
parameter space for other maps and topologies are presented in the Supplemental Material
\cite{SuppMat}.

Specifically, in Fig.~\ref{fig_4} we see how the region where $\min(\Delta) = 0$ for fixed
$N$ in the $(\epsilon,\,p)$ space changes depending on the units dynamical behavior. In
particular, we note that with the exception of a robust window located for $\epsilon\in
(0.02,0.10)$ and $p < 0.5$ where fully incoherent behavior is found [dark region in
Fig.~\ref{fig_4}{\bf (a)} and {\bf (b)}], coupled chaotic logistic maps have periodic
windows [light region in Fig.~\ref{fig_4}{\bf (b)}] and synchronized behavior [triangular
region in the upper corner of Fig.~\ref{fig_4}{\bf (a)} and {\bf (b)}] which make the
inference impossible. On the other hand, no coherent behavior is found for the circle maps
in the same $(\epsilon,\,p)$ space [Fig.~\ref{fig_4}{\bf (c)} and {\bf (d)}]. Thus, the
region where perfect inference is possible (white circles), is mainly limited by the amount
of data available (disregarding $\epsilon \sim 0$). \emph{The $\epsilon$-robustness of the
CC or MI results, depends on the dynamic of the units composing the system and the topology
$(p)$. Although, these results are reliable and the methods are effective if dynamical
coherence is avoided and sufficient data is available}. In order to retrieve similar regions
of perfect inference for larger ($N$) networks, we find that larger data sets are needed
(See Supplemental Material \cite{SuppMat}).

A similar conclusion is drawn when analyzing the effect of increasing the networks
size $N$ but keeping the time-series length $T$ fixed. Namely, we note that the region in
the $(\epsilon,\,p)$ space that perfect inference is possible diminishes as $N$ is increased
if $T$ is kept fixed (see Supplemental Material \cite{SuppMat}). Thus, it is expected that
an analysis which uses $N/T^\alpha$ constant, with $\alpha > 1$ as the control parameter,
maintains the regions of perfect inference in the $(\epsilon,\,p)$ invariant.

 \section{Conclusions}
To conclude, we have shown that the Cross-Correlation coefficient (CC, calculated in absolute
value) and the Mutual Information (MI, calculated from ordinal patterns) are able to infer,
without errors, the underlying topology of different coupled discrete maps when there is an
abrupt change in the ordered set of their magnitudes. We showed that, while both methods
require weakly coupled units (avoiding the presence of global patterns, or strong
desynchronization) for the abrupt change to exist, the MI is in general more robust and
reliable. To the best of our knowledge, reliable reconstruction of network topologies without
errors from time-series measurements of discrete-time dynamical units has not been previously
obtained.

Various fields where complex networks of interactions are often inferred via a CC or MI
statistical similarity analysis of observed time-series can benefit from our results. A
careful consideration of the shape of the distribution of similarity values could allow for
selecting optimal thresholds for the inference of direct links, as opposite to the often
employed methods based on quantiles or on deviations from surrogate data.

 \section{Acknowledgements}
Authors N.R., E.B.-M., C.G., and M.S.B. acknowledge the Scottish Universities Physics
Alliance (SUPA). E. B.-M. and M.S.B. also acknowledge the Engineering and Physical Science
Research Council (EPSRC) project Ref. EP/I032 606/1. A.C.M and C.M acknowledge the LINC
project (FP7-PEOPLE-2011-ITN, grant No. 289447). A.C.M also aknowledges PEDECIBA and CSIC
(Uruguay). C.M also acknowledges grant FIS2012-37655-C02-01 from the Spanish MCI, grant
2009 SGR 1168, and the ICREA Academia programme from the Generalitat de Catalunya.



\end{document}